\documentclass[12pt, preprint]{aastex}
\usepackage{amsmath}
\usepackage{graphicx}
\usepackage{natbib}
\usepackage{ulem}

\usepackage[colorlinks=true,citecolor=blue,breaklinks=true,linktocpage=true]{hyperref}
\bibpunct{(}{)}{;}{a}{}{,} 
\usepackage{xspace}
\usepackage{hyperref}

\usepackage{multirow}

\def\vec#1{\ensuremath{\mathbf{#1}}}

\usepackage{color}

\newcommand{\omits}[1]{}

\shorttitle{Slow surface sausage modes in photospheric waveguides}
\shortauthors{Chen et al.}

\begin{document}


\title{Damping of slow surface sausage modes in photospheric waveguides}

\author{Shao-Xia Chen\altaffilmark{1}}
\author{Bo Li\altaffilmark{1}}
    \email{bbl@sdu.edu.cn}
\author{Mijie Shi\altaffilmark{1}}
\and
\author{Hui Yu\altaffilmark{1}}

\altaffiltext{1}{Shandong Provincial Key Laboratory of Optical Astronomy and Solar-Terrestrial Environment,
   Institute of Space Sciences, Shandong University, Weihai 264209, China}

\begin{abstract}
There has been considerable interest in sausage modes in photospheric waveguides like
    pores and sunspots, and slow surface sausage modes (SSSMs) have been suggested to damp
    sufficiently rapidly to account for chromospheric heating.
Working in the framework of linear resistive magnetohydrodynamics, we examine how efficient
    electric resistivity and resonant absorption in the cusp continuum can be for damping SSSMs
    in a photospheric waveguide with equilibrium parameters compatible with recent measurements of a photospheric pore.
For SSSMs with the measured wavelength, we find that the damping rate due to the cusp resonance
    is substantially less strong than theoretically expected with the thin-boundary approximation.
The damping-time-to-period ratio ($\tau/P$) we derive for standing modes,
    equivalent to the damping-length-to-wavelength ratio for propagating modes given the extremely weak dispersion,
    can reach only $\sim 180$.
However, the accepted values for electric resistivity ($\eta$) correspond to a regime
    where both the cusp resonance and resistivity play a role.
The values for $\tau/P$ attained at the largest allowed $\eta$ may reach $\sim 30$.
We conclude that electric resistivity can be considerably more efficient than the cusp resonance
    for damping SSSMs in the pore in question, and it needs to be incorporated into future studies
    on the damping of SSSMs in photospheric waveguides in general.

\end{abstract}

\section{INTRODUCTION}
\label{sec_intro}
Identifying {magnetohydrodynamic (MHD)} waves in the structured solar atmosphere
    is vital from the perspectives of both solar magneto-seismology
    \citep[SMS, see recent reviews by, e.g.,][]{2005LRSP....2....3N,2012RSPTA.370.3193D,2016SSRv..200...75N}
    and atmospheric heating \citep[see e.g.,][for recent reviews]{2015RSPTA.37340261A, 2015RSPTA.37340269D}.
While coronal waves and oscillations remain a focus
    \citep[][to name only a few]{1999ApJ...520..880A,1999Sci...285..862N,2002ApJ...574L.101W,
    2008A&A...482L...9O,2008A&A...489L..49E, 2009ApJ...698..397V,2011ApJ...736..102A,2016A&A...593A..53P},
    various MHD modes have also been identified
    in a considerable number of structures in the lower solar atmosphere
    (e.g., \citeauthor{2007Sci...318.1574D}~\citeyear{2007Sci...318.1574D},
      \citeauthor{2009Sci...323.1582J}~\citeyear{2009Sci...323.1582J},
      \citeauthor{2009ApJ...705L.217H}~\citeyear{2009Sci...323.1582J},
      \citeauthor{2012NatCo...3E1315M}~\citeyear{2012NatCo...3E1315M},
      \citeauthor{2014Sci...346A.315T}~\citeyear{2014Sci...346A.315T},
      \citeauthor{2017ApJ...842...59J}~\citeyear{2017ApJ...842...59J},
      \citeauthor{2018ApJ...856L..16W}~\citeyear{2018ApJ...856L..16W};
      see also the review by \citeauthor{2012LRSP....9....2A}~\citeyear{2012LRSP....9....2A}
         and references therein).
In particular, photospheric pores and sunspots were shown to host a variety of
    waves.
Thanks to modern high-cadence and high-spatial-resolution
    instruments like the Rapid Oscillations in the Solar Atmosphere
    \citep[ROSA,][]{2010SoPh..261..363J},
    these waves were ascertained, on the bases of parity and axial phase speed,
    to be either fast or slow sausage waves~\citep{2008IAUS..247..351D, 2011ApJ...729L..18M,
    2014A&A...563A..12D,2015A&A...579A..73M,2015ApJ...806..132G,2016ApJ...817...44F}.
More recently, the measurements of the spatial distribution of wave power
    enabled these waves to be further grouped into surface and body modes
    \citep{2018ApJ...857...28K}.

Among the above-mentioned studies on waves in photospheric structures, 
    of particular interest are
    the recent multi-wavelength measurements of slow surface sausage modes (SSSMs)
    in a magnetic pore, where propagating SSSMs were shown to damp
    over a lengthscale as short as a quarter of the wavelength so as
    to be important in heating
    the chromosphere \citep[][hereafter G15]{2015ApJ...806..132G}.
Working with the thin-boundary (TB) approximation,
    \citeauthor{2017A&A...602A.108Y}~(\citeyear{2017A&A...602A.108Y}, hereafter Y17;
    also \citeauthor{2017ApJ...850...44Y}~\citeyear{2017ApJ...850...44Y})
    showed that the resonant absorption of SSSMs in the cusp continuum may play a substantial role
    in their damping, with the damping-time-to-period ratio ($\tau/P$)
    attaining down to $\sim 10$.
Note that while this damping was obtained for standing modes (i.e., real axial wavenumbers),
    the damping-length-to-wavelength ratio ($L_{\rm D}/\lambda$) is nearly identical to $\tau/P$
    because of the extremely weak dispersion of SSSMs.
Note further that even though the theoretical framework in the TB limit was well-established
    \citep[see the review by][and references therein]{2011SSRv..158..289G},
    Y17 were the first to show that the cusp resonance of SSSMs is stronger than thought,
    even though the values of $L_{\rm D}/\lambda$ are not sufficient to account for
    the measured heavy damping.

The purpose of this study is to further examine the damping of SSSMs in photospheric waveguides
    by solving the linearized resistive MHD equations.
By so doing, we will address how efficient the damping due to the cusp resonance is in a more general sense
    where the TB approximation is not invoked.
In addition, we will also be able to assess the role of electric resistivity ($\eta$) for damping SSSMs.
This latter point is important in that the magnetic Reynolds numbers ($R_{\rm m}$) in pores and sunspots
    may be orders-of-magnitude smaller than typical coronal values ($\sim 10^{12}-10^{14}$).
Take the pore examined by G15, for which
    the radius $R\sim 1500$~km and the internal Alfv\'en speed $v_{\rm Ai} \sim 12$~km~s$^{-1}$.
For electric conductivity ($\sigma = 1/\eta$), we quote values in the range between $1.3$
    and $420$~S~m$^{-1}$, which were in fact derived
    by \citet{1983SoPh...84...45K} for a model sunspot umbra in the temperature range of $3500$
    to $\sim 6400$~K.
One finds a range between $3\times 10^4$ and $10^7$ for $R_{\rm m}$ as defined by $\mu_0 \sigma R v_{\rm Ai}$,
    where $\mu_0$ is the magnetic permeability in free space.
\footnote{The range of electric conductivity is also compatible with, e.g.,
    \citet[][Fig.~1]{1993ASPC...46..465W} and \citet[][Fig.~1]{2012ApJ...747...87K},
    see also \citet{2016ApJ...832..195N}.
   Note that the resistivity we address corresponds to the classical Ohmic one using
   the terminology of \citet{2012ApJ...747...87K}.}
As is well-known, resonant absorption is essentially an ideal MHD process whereby the attenuation of
    the collective modes is due to their energy being transferred to localized waves (either Alfv\'en or slow or both),
    even though these localized waves are eventually dissipated.
From the perspective of resistive MHD, the associated damping rates can be found from
    the eigen-frequencies $\omega$ at sufficiently large $R_{\rm m}$ such that $\omega$ is independent of $R_{\rm m}$
    \citep[see e.g.,][for kink waves resonantly coupled to the Alfv\'en continuum]
    {1991PhRvL..66.2871P,2004ApJ...606.1223V,2006ApJ...642..533T,2016SoPh..291..877G}.
However, it is not known a priori whether the above-quoted $R_{\rm m}$ is sufficiently large.
So in short, we would like to address two questions:
One, what happens to the damping rate due to the cusp resonance if the waveguide boundary is not thin?
Second, how does the damping efficiency due to direct resistive dissipation
    compare with that associated with the cusp resonance?
Our problem formulation is given in Section~\ref{sec_model}
    and the numerical results are presented in Section~\ref{sec_para_study}.
Section~\ref{sec_conclusion} summarizes this study and ends with some concluding remarks.

\section{MATHEMATICAL FORMULATION}
\label{sec_model}

\begin{figure}
\centering
 \includegraphics[width=.6\columnwidth]{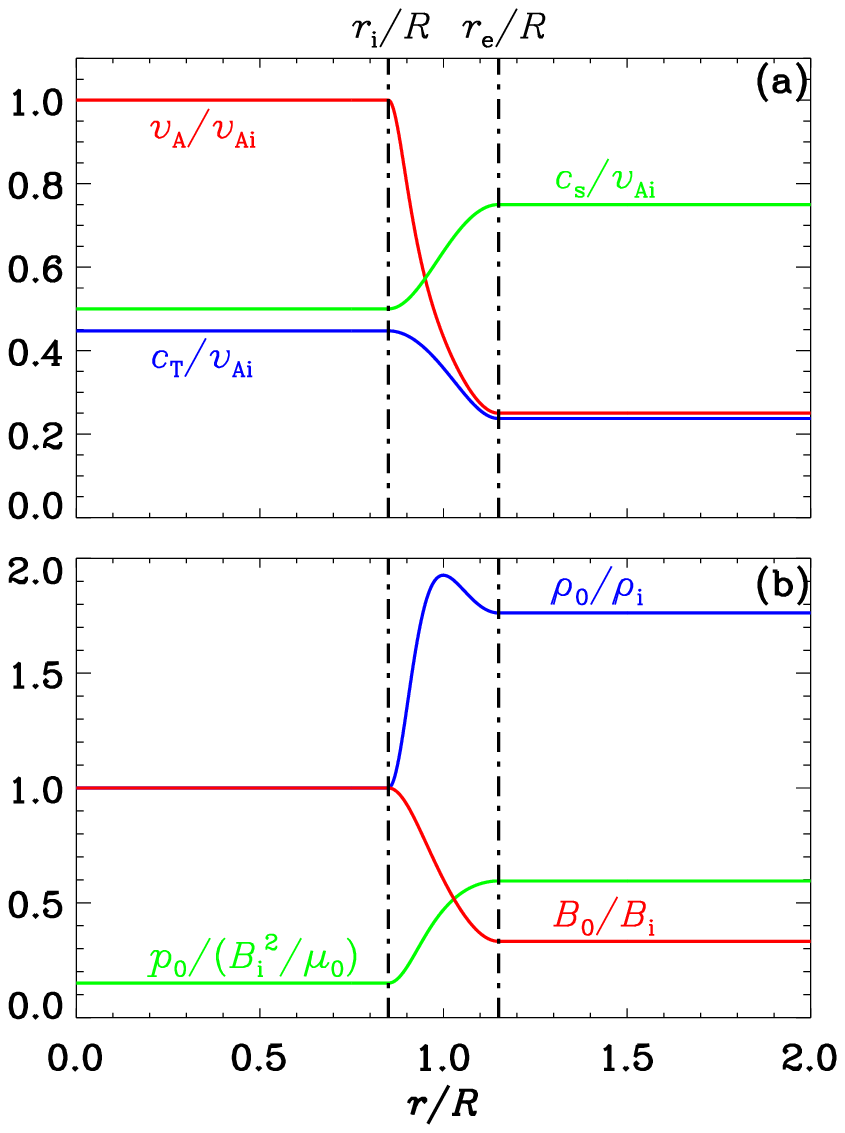}
 \caption{Transverse profiles for the equilibrium parameters of a photospheric waveguide
     representative of a pore.
 {Both the characteristic speeds (the upper panel) 
     and equilibrium fluid parameters (lower) are shown.
 Characteristic of these profiles is that 
     a continuous transition layer (TL) connects a uniform cord and a uniform exterior}.
 This TL is located between $r_{\rm i} = R-l/2$ and $r_{\rm e} = R+l/2$, where
     $R$ is the mean waveguide radius and $l$ the TL width.
 Here $l/R$ is arbitrarily chosen to be $0.3$ for illustration purposes.      
 }
 \label{fig_EQprofile}
\end{figure}

With magnetic pores in mind, we model photospheric waveguides as static, straight magnetic cylinders
     aligned with the equilibrium magnetic field $\vec{B}_0 = B_0 \hat{z}$
     in a cylindrical coordinate system $(r, \theta, z)$.
We neglect gravity throughout, and hence the equilibrium parameters (denoted by subscript $0$)
     depend only on $r$.
Let $\rho$, $p$, and $\vec{v}$ denote mass density, thermal pressure, and velocity, respectively.
The adiabatic sound speed $c_{\rm s}$, Alfv\'en speed $v_{\rm A}$,
     and cusp (or tube) speed $c_{\rm T}$ are defined by the equilibrium values as
\begin{eqnarray}
\label{eq_def_cs_va_cT}
\displaystyle
c_{\rm s}^2 = \frac{\gamma p_0}{\rho_0}~, \hspace{0.2cm}
v_{\rm A}^2 = \frac{B_0^2}{\mu_0 \rho_0}~,\hspace{0.2cm}
c_{\rm T}^2 = \frac{c_{\rm s}^2 v_{\rm A}^2}{c_{\rm s}^2 + v_{\rm A}^2}~,
\end{eqnarray}
   where $\gamma=5/3$ is the adiabatic index.
The transverse distribution of the equilibrium parameters is assumed to
    comprise a uniform cord (denoted by subscript ${\rm i}$),
    a uniform external medium (subscript ${\rm e}$),
    and a continuous transition layer (TL) connecting the two.
To realize this, we assume that $c_{\rm T}^2$ and $c_{\rm s}^2$ in the TL
    are given by
\begin{eqnarray}
\displaystyle
{\cal E}_{\rm TL}(r)=
    \frac{{\cal E}_{\rm i}+{\cal E}_{\rm e}}{2}
   -\frac{{\cal E}_{\rm i}-{\cal E}_{\rm e}}{2}
     \sin\frac{\pi(r-R)}{l}~,				
\label{eq_profile_cs_ct}
\end{eqnarray}
    which applies for $R-l/2 \equiv r_{\rm i} \le r \le r_{\rm e} \equiv R+l/2$.
Here ${\cal E}$ represents $c_{\rm T}^2$ and $c_{\rm s}^2$,
    $R$ denotes the mean radius, and $l$ the TL width.
Identical to \citet{1983SoPh...88..179E},
    we take $[c_{\rm si}, c_{\rm se}, v_{\rm Ae}] = [0.5, 0.75, 0.25] v_{\rm Ai}$,
    resulting in $[c_{\rm Ti}, c_{\rm Te}] = [0.4472, 0.2372] v_{\rm Ai}$.
If taking $v_{\rm Ai} = 12$~km~s$^{-1}$,
    then one finds that $[c_{\rm si}, c_{\rm se}, v_{\rm Ae}]= [6, 9, 3]$~km~s$^{-1}$,
    largely compatible with G15 for the height range of $[0, 1000]$~km (see their Fig. 4).
Note that if we see $R$, $\rho_{\rm i}$ and $v_{\rm Ai}$ as normalizing constants,
    and once $l/R$ is specified,
    then the equilibrium profile is complete as a result of
    transverse force balance, namely,
\begin{equation}
\displaystyle
   p_0 + \frac{B_0^2}{2\mu_0}
     = \rho_0\left(\frac{c^2_{\rm s}}{\gamma}+\frac{v^2_{\rm A}}{2} \right)
     = \mbox{const}~.
\label{eq_force_balance}
\end{equation}
From the set of independent constants $[R, \rho_{\rm i}, v_{\rm Ai}]$,
    we derive the constant for normalizing magnetic field strength as
    $\sqrt{\mu_0 \rho_{\rm i} v_{\rm Ai}^2}$, which equals $B_{\rm i}$ by definition.
Then the constants to normalize time and thermal pressure are
    $R/v_{\rm Ai}$ and $\rho_{\rm i} v_{\rm Ai}^2 = B^2_{\rm i}/\mu_0$,
    respectively.
Figure~\ref{fig_EQprofile} presents the transverse distributions of the relevant equilibrium
    parameters with $l/R$ arbitrarily chosen to be $0.3$ for illustration purposes.
{Both the characteristic speeds (Figure~\ref{fig_EQprofile}a)
    and the equilibrium fluid parameters (Figure~\ref{fig_EQprofile}b)
    are shown. 
With $c_{\rm T}^2$ and $c_{\rm s}^2$ specified by Equation~\eqref{eq_profile_cs_ct},
    all the profiles show a monotonical dependence on $r$ with the exception of
    the equilibrium density $\rho_0$.
If we prescribe, say, a linear dependence on $r$ of 
    the equilibrium density and pressure profiles, the cusp speed $c_{\rm T}$
    may possess a nonmonotonical $r$-dependence and hence 
    SSSMs may experience multiple resonances in the TL
    (see \citeauthor{2017ApJ...850...44Y}~\citeyear{2017ApJ...850...44Y}
    for an example). 
We choose the present procedure for specifying the equilibrium profiles to avoid 
    this further complication.}

Let $\vec{v}_1$, $p_1$ and $\vec{B}_1$ denote the perturbations to
    velocity, pressure, and magnetic field, respectively.
The small-amplitude perturbations to the equilibrium are then governed by
    the following set of linearized, resistive MHD equations
\begin{eqnarray}
&&  \displaystyle
   \rho_0 \frac{\partial \vec{v}_{1}}{\partial t}
     = -\nabla p_1 +\frac{(\nabla\times\vec{B}_0)\times\vec{B}_1}{\mu_0}
                   +\frac{(\nabla\times\vec{B}_1)\times\vec{B}_0}{\mu_0}~,
                   \label{eq_linMHD_momen} \\
&&  \displaystyle
   \frac{\partial \vec{B}_1}{\partial t}
     = \nabla\times\left(\vec{v}_1\times\vec{B}_0-\frac{\eta}{\mu_0}\nabla\times\vec{B}_1
                   \right)~, \label{eq_linMHD_Farad} \\
&&  \displaystyle
   \frac{\partial p_1}{\partial t}
     = -\vec{v}_1\cdot\nabla p_0 -\gamma p_0 \nabla\cdot\vec{v}_1
       + 2\frac{(\gamma-1)\eta}{\mu_0^2}
            \left(\nabla\times\vec{B}_1\right)\cdot\left(\nabla\times\vec{B}_0\right)~,
            \label{eq_linMHD_pres}
\end{eqnarray}
    which are in dimensional form.
The electric resistivity $\eta$ is assumed to be uniform for simplicity.
With our normalizing constants, the dimensionless form of Equations~\eqref{eq_linMHD_momen} to \eqref{eq_linMHD_pres}
    differs from their dimensional form only in two aspects.
First, $\mu_0$ disappears in Equation~\eqref{eq_linMHD_momen}.
Second, $\eta/\mu_0$ and $\eta/\mu_0^2$ in Equations~\eqref{eq_linMHD_Farad}
    and \eqref{eq_linMHD_pres} are both replaced with $1/R_{\rm m}$,
    where $R_{\rm m} = \mu_0 R v_{\rm Ai}/\eta$ is the magnetic Reynolds number.
\footnote{
Strictly speaking, the equilibrium cannot stay static with a spatially varying $B_0$ given
    a finite resistivity.
An equilibrium flow $\vec{v}_0 = v_0 \hat{r}$ is needed to counteract magnetic diffusion,
    where $\hat{r}$ is the unit vector in the $r$-direction.
However, balancing $\vec{v}_0 \times \vec{B}_0$ with $\eta \nabla\times\vec{B}_0/\mu_0$ yields that
\begin{equation}
\displaystyle
    \frac{v_0}{v_{\rm Ai}}  = \frac{1}{R_{\rm m}}\frac{{\rm d} \ln B_0}{{\rm d} r/R}~.
\end{equation}
With $l/R \ge 0.01$ and $R_{\rm m} \ge 10^4$ as we adopt,
    $|v_0|$ is $\lesssim 1.5\times 10^{-2} v_{\rm Ai}$ or $180$~m~s$^{-1}$ in absolute terms.
}

Focusing on sausage modes, we adopt the following ansatz for any perturbation
\begin{eqnarray}
\label{eq_Fourier_ansatz}
    f_1(r,z;t)={\rm Re}\left\{\tilde{f}(r)\exp\left[-i\left(\omega t-kz\right)\right]\right\}~,
\end{eqnarray}
    where $\omega$ is the complex-valued angular frequency
    and $k$ the real-valued axial wavenumber.
Only the equations governing $\tilde{v}_r$, $\tilde{v}_z$, $\tilde{B}_r$, $\tilde{B}_z$
    and $\tilde{p}$ survive.
{To be specific, they read 
\begin{eqnarray}
\omega\tilde{v}_r & = & \displaystyle
     -\frac{B_0}{\rho_0}\left(k\tilde{B}_r +i\frac{{\rm d}\tilde{B}_z}{{\rm d}r}\right)
     -\frac{i \tilde{B}_z}{\rho_0}\frac{{\rm d}B_0}{{\rm d}r}
     -\frac{i}{\rho_0}\frac{{\rm d}\tilde{p}}{{\rm d}r}~,  \label{eq_Fourier_vr}\\
\omega\tilde{v}_z & = & \displaystyle
      \frac{i \tilde{B}_r}{\rho_0}\frac{{\rm d}B_0}{{\rm d}r}
     +\frac{k\tilde{p}}{\rho_0}~, 			   \label{eq_Fourier_vz}	\\
\omega\tilde{B}_r & = & \displaystyle
     -kB_0\tilde{v}_r +
     \frac{1}{R_{\rm m}}\left(i\frac{{\rm d^2}\tilde{b}_r}{{\rm d}r^2}
         +\frac{i}{r}\frac{{\rm d}\tilde{b}_r}{{\rm d}r}-ik^2\tilde{b}_r
         -\frac{i\tilde{b}_r}{r^2}
     \right)~, 						   \label{eq_Fourier_Br} \\
\omega\tilde{B}_z & = & \displaystyle
    -i\tilde{v}_r\frac{{\rm d}B_0}{{\rm d}r}
    -B_0\left(i\frac{{\rm d}\tilde{v}_r}{{\rm d}r}+i\frac{\tilde{v}_r}{r}\right)
    +\frac{1}{R_{\rm m}}\left(i\frac{{\rm d^2}\tilde{b}_z}{{\rm d}r^2}
         +\frac{i}{r}\frac{{\rm d}\tilde{b}_z}{{\rm d}r}-ik^2\tilde{b}_z
     \right)~,						   \label{eq_Fourier_Bz}	\\
\omega\tilde{p} & = & \displaystyle
    -c^2_{\rm s}\rho_0
    \left(i\frac{{\rm d}\tilde{v}_r}{{\rm d}r}
         +i\frac{\tilde{v}_r}{r}-k\tilde{v}_z
       \right)
    -i\tilde{v}_r\frac{{\rm d}p_0}{{\rm d}r} 
    +\frac{2\left(\gamma-1\right)}{R_{\rm m}}\frac{{\rm d}B_0}{{\rm d}r} 
       \left(k\tilde{B}_r+i\frac{{\rm d}\tilde{B}_z}{{\rm d}r}\right)~,  \label{eq_Fourier_p}
\end{eqnarray}
    which are in dimensionless form.
}
These equations constitute a standard eigen-value problem (EVP) when supplemented with proper boundary conditions.
At the cylinder axis ($r=0$), the parity of sausage modes dictates that
    $\tilde{v}_r = \tilde{B}_r = 0$ and the $r$-derivatives of the rest vanish.
Infinitely far from the cylinder, all dependent variables are zero.
Taking $[l/R, kR, R_{\rm m}]$ as free parameters, we then solve the EVP
    {with the code PDE2D}~\citep{1988Sewell_PDE2D}, which was first introduced to the solar context
    by \citet{2006ApJ...642..533T}.
A nonuniform grid is set up for $r$ between $0$ and $r_{\rm M}$, 
    {where $r_{\rm M}$ is the outer boundary.
A substantial} 
    number of grid points are deployed in the TL to resolve
    the possible oscillatory behavior therein.
{We place $r_{\rm M}$ at $50~R$,
    and make} sure that further increasing $r_{\rm M}$ does not introduce any appreciable change
    to the computed eigen-frequencies ($\omega$).
In short, $\omega$ is then formally expressed as
\begin{eqnarray}
    \frac{\omega R}{v_{\rm Ai}} = {\cal G}\left(kR, \frac{l}{R}, R_{\rm m}\right)~.
\label{eq_omega_formal}
\end{eqnarray}
The wave period $P$ and damping time $\tau$ are given by
    $P=2 \pi/\omega_{\rm R}$ and $\tau = 1/|\omega_{\rm I}|$, respectively.
We also define the axial phase speed $v_{\rm ph}$ as $\omega_{\rm R}/k$,
    and the axial group speed $v_{\rm gr}$ as ${\rm d}\omega_{\rm R}/{\rm d}k$.
Here we use the shorthand notations $\omega_{\rm R} = \rm{Re}(\omega)$
    and $\omega_{\rm I} = \rm{Im}(\omega)$.

The damping rates associated with the cusp resonance are
    also computed in the TB limit ($l/R \ll 1$).
Analytical treatments in this limit yield a
    dispersion relation (DR) for surface sausage modes
    as given by Equation~(27) in Y17, where resistivity is not explicitly involved.
Note that the location ($r_{\rm c}$) where the cusp resonance takes place
    is not known beforehand.
We therefore solve the pertinent DR by iteration.
For any given set of $[l/R, kR]$, we solve the DR for $\omega$ with an initial guess of $r_{\rm c}$,
    and then locate $r_{\rm c}$ with the computed $\omega_{\rm R}$ such that $\omega_{\rm R}/k$
    equals $c_{\rm T}$.
Correcting the guess for $r_{\rm c}$, we repeat this process until convergence is met.
These TB eigen-frequencies will be compared with those at sufficiently large $R_{\rm m}$
    from our resistive computations, where $l/R$ is allowed to be arbitrary.
\footnote{
If we are interested only in the cusp resonance for transition layers not that thin,
    then we can adopt the Frobenius approach
    by \citet{2013ApJ...777..158S} who worked exclusively in ideal MHD.
This approach yields an analytical DR whose validity is not restricted
    to the TB limit.
However, we adopt the resistive MHD approach because we are interested also in whether resistivity
    plays a role in wave damping.
}

\section{NUMERICAL RESULTS}
\label{sec_para_study}

\begin{figure}
\centering
 \includegraphics[width=1.\columnwidth]{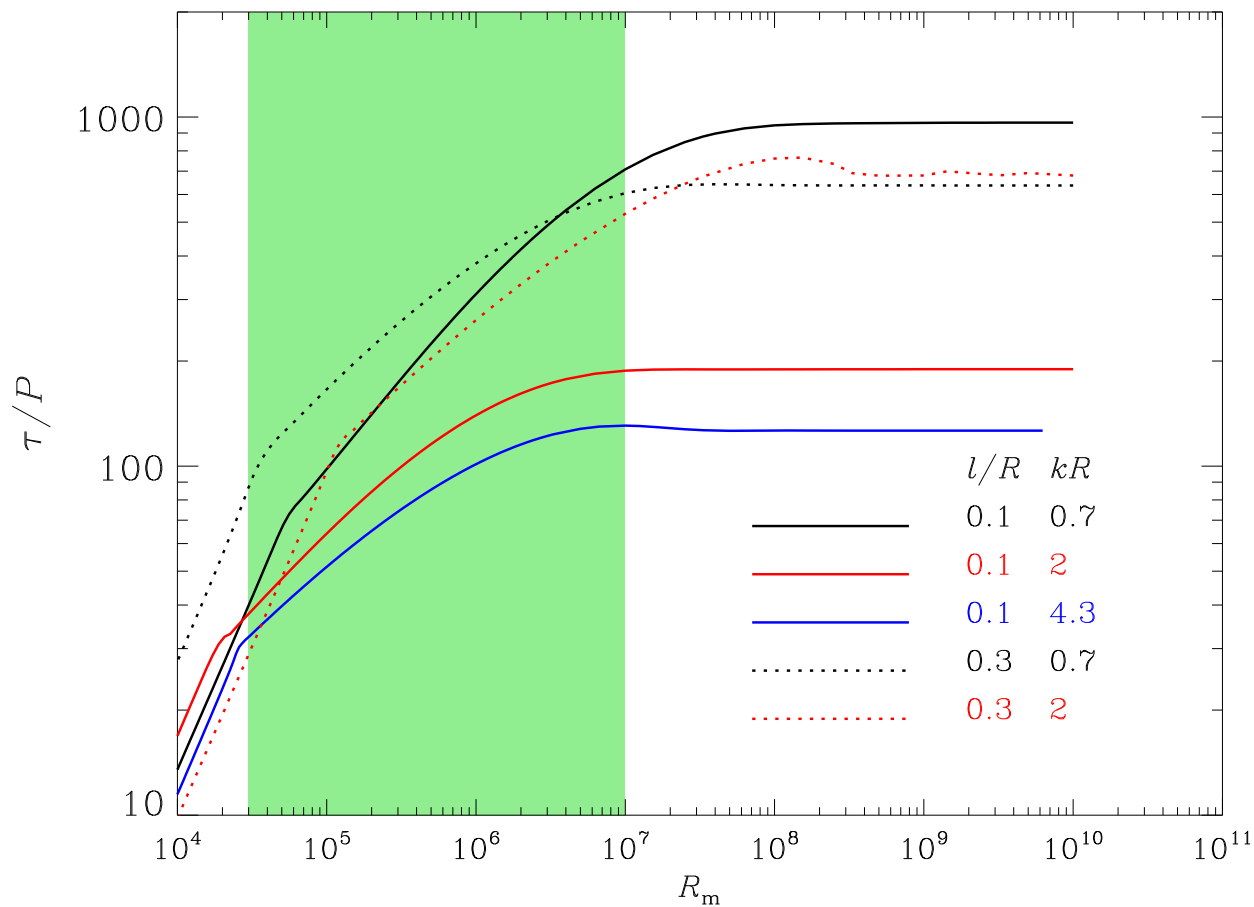}
 \caption{Dependence on the magnetic Reynolds number ($R_{\rm m}$) of
    the damping-time-to-period ratio ($\tau/P$) of
    slow surface sausage modes in a photospheric waveguide
     representative of a pore.
 A number of combinations of the transition layer width $l$
     and axial wavenumber $k$ are examined as labeled.
 The area shaded green corresponds to the range of $R_{\rm m}$
     derived from accepted values of electric conductivity.
 See text for details.     
 }
 \label{fig_dep_resis}
\end{figure}

To start, Figure~\ref{fig_dep_resis} shows the dependence on $R_{\rm m}$ of the damping-time-to-period ratio
    $\tau/P$ of slow surface sausage modes (SSSMs)
    for a number of combinations of $[l/R, kR]$ as labeled.
We will come back to this figure later, for now it suffices to note that
    $\tau/P$ tends to increase with $R_{\rm m}$ and eventually levels off.
These saturation values are then taken as the values of $\omega$
    pertinent to resonant absorption for each pair of $[l/R, kR]$.
By saturation, we mean that the change in $\tau/P$ does not exceed $1.5\%$ for
    $R_{\rm m}$ between $10^9$ and $10^{10}$.
It turns out that saturation is not guaranteed when either $kR$ or $l/R$ is sufficiently large
    even for $R_{\rm m}$ as large as $10^{10}$
    (such a hint can be seen from the red dotted curve pertaining to
    $[l/R, kR] = [0.3, 2]$).
Note that here we followed a rather stringent definition of ``saturation''.
Examining the red dotted curve in Figure~\ref{fig_dep_resis}, one sees that 
    $\tau/P$ depends only rather weakly on $R_{\rm m}$ when 
    $R_{\rm m} \gtrsim 3\times 10^8$.
However, that $\tau/P$ shows a clear dependence on resistivity for $R_{\rm m}$ as large
    as $\sim 10^8$ is quite different from what happens to standing kink modes resonantly coupled 
    to the Alfv\'en continuum in typical coronal loops.
In the latter case, typically $\tau/P$ becomes independent of resistivity when $R_{\rm m}$
    exceeds $\sim 10^{5}$ 
    (see e.g., \citeauthor{2006ApJ...642..533T}~\citeyear{2006ApJ...642..533T}, Fig.~2;
    \citeauthor{2016SoPh..291..877G}~\citeyear{2016SoPh..291..877G}, Fig.~9)
    or $10^{6}$~\citep[e.g.,][Fig.~3]{2004ApJ...606.1223V}.
We tentatively attribute the differences in the dependence of $\tau/P$ on resistivity 
    to two factors.
First, the equilibrium parameters in this study are very different from the coronal cases.
This is particularly true for the ordering of the characteristic speeds in and outside the waveguide.
Second, the axial wavenumbers themselves are quite different.
To be more specific, $kR = 2$ for the red dotted curve in Figure~\ref{fig_dep_resis},
    whereas $kR$ is typically taken to be $\lesssim 0.1\pi$ in studies on
    standing kink modes in coronal loops.
In what follows, we show only the saturation values.

\begin{figure}
\centering
 \includegraphics[width=1.\columnwidth]{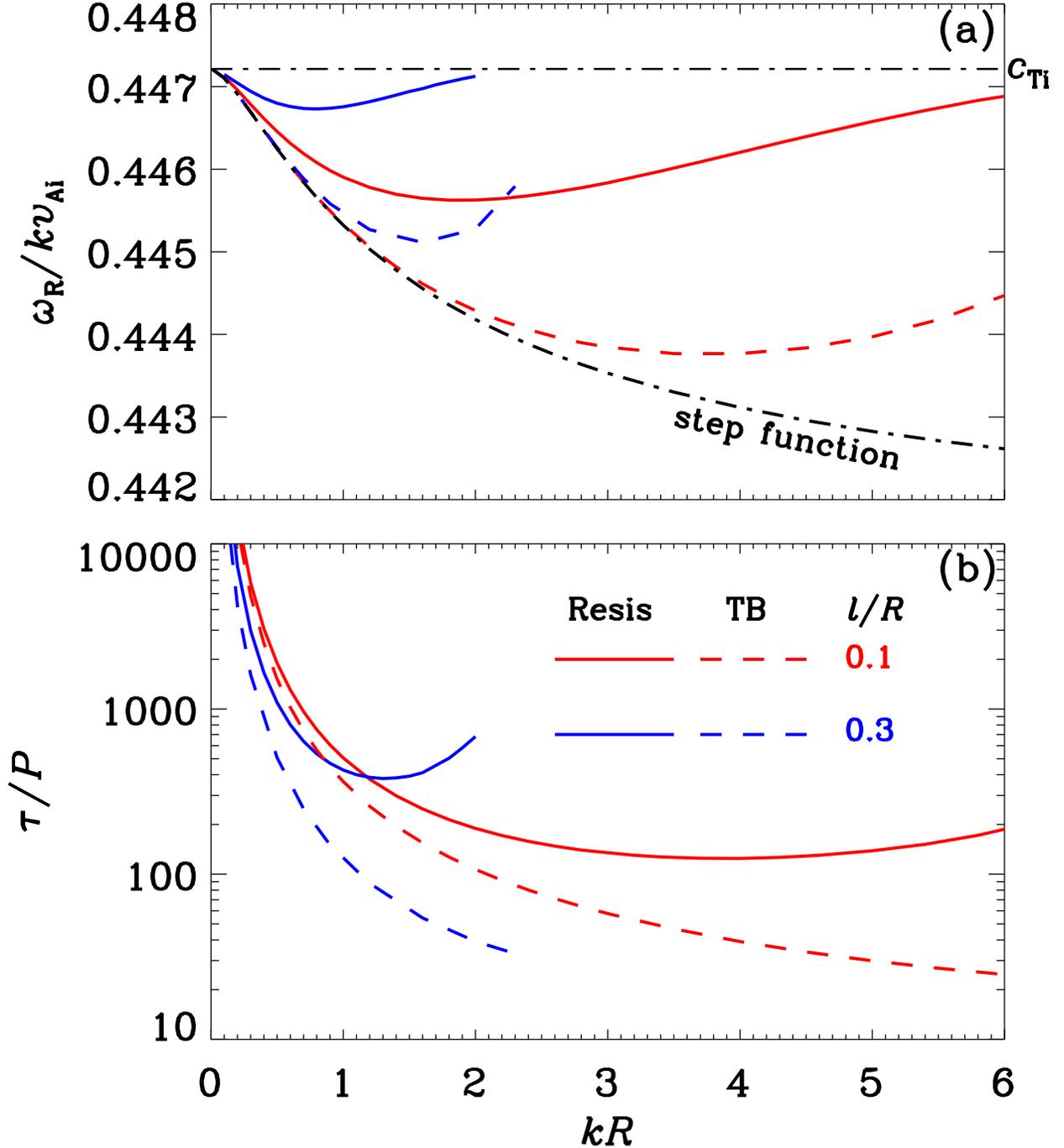}
 \caption{Dispersive properties of slow surface sausage modes resonantly coupled to the cusp continuum,
   shown by the dependence on the axial wavenumber ($k$) of
   (a) the axial phase speed ($\omega_{\rm R}/(k v_{\rm Ai})$) and
   (b) the damping-time-to-period ratio ($\tau/P$).
 Two values of the transition layer width $l$ (in units of the mean waveguide radius $R$)
   are presented by the lines with different colors as labeled.
 In addition to the results from our resistive computations (the solid curves),
    those from the thin-boundary (TB) computations are also shown (dashed).
 The black dash-dotted curve in panel (a) corresponds to the case where $l/R = 0$, namely
    the equilibrium parameters are transversely structured as a step function.
 }
 \label{fig_dep_k}
\end{figure}

Figure~\ref{fig_dep_k} shows how the axial phase speed $v_{\rm ph}=\omega_{\rm R}/k$
    and damping-time-to-period ratio $\tau/P$ depend
    on the axial wavenumber $k$ for a number of $l/R$ as labeled.
In addition to the results from the resistive computations (the solid lines),
    we also show the results found by solving the TB limit
    (dashed).
The dash-dotted curve in Figure~\ref{fig_dep_k}a shows what happens when
    $l/R = 0$.
This corresponds to the case where the equilibrium parameters are transversely structured
    in a step-function fashion, and hence the label ``step function''.
{Figure~\ref{fig_dep_k}a shows that}
    $\omega_{\rm R}/k$ for $l/R\ne 0$ does not show
    a monotonical dependence on $k$ as in the step-function case.
Rather, $\omega_{\rm R}/k$ decreases with $k$ from the internal tube speed $c_{\rm Ti}$,
    attains a local minimum, and then increases toward $c_{\rm Ti}$ again.
This behavior is seen not only for the resistive results (the solid curves),
    but also for the TB ones (dashed).
The latter behavior requires some explanation.
In the case of weak damping, one usually proceeds by first solving the real part of the DR in the TB limit
    for $\omega_{\rm R}$, which is then used to find $r_{\rm c}$
    such that $\omega_{\rm I}$ can be found~(see Y17 and references therein).
With such an analytical approach, the TB results in Figure~\ref{fig_dep_k}a will agree with
    the step-function case exactly.
However, the nominal real part of the DR actually involves the complex-valued $\omega$
    (see Equation~30 in Y17), and this turns out to have a subtle effect
    on the solutions.
Indeed, \citet{2017ApJ...850...44Y} also went beyond the two-step procedure, and
    showed that the full numerical solution to the DR in the TB limit
    yields a nonmonotonical dependence of $\omega_{\rm R}/k$ on $k$ as well
    (Figure 5 therein).
Note, however, that the equilibrium profiles adopted in \citet{2017ApJ...850...44Y}
    are different from ours and a direct comparison
    is not straightforward.
What the dashed lines in our Figure~\ref{fig_dep_k}a show is that the nonmonotonical behavior
    persists for our choice of the equilibrium profile in the TL.

Figure~\ref{fig_dep_k}b is more relevant for examining how efficient
    resonant absorption may be to attenuate SSSMs.
Examine the solid lines first, which pertain to the resistive computations.
One sees that $\tau/P$ decreases first with $kR$ but eventually increases again, and therefore
    for a chosen $l/R$ there is an optimal axial wavenumber
    that maximizes wave attenuation.
The increase in $\tau/P$ at large $kR$ occurs in conjunction with the tendency for the axial
    phase speed $\omega_{\rm R}/k$ to approach $c_{\rm Ti}$.
Note that the $\tau/P-kR$ curve for a given $l/R$ terminates when $kR$ is sufficiently large.
This does not mean that resonant SSSMs
    are prohibited for axial wavenumbers beyond some critical value.
Rather, this comes from the issue that the eigen-frequencies still show some weak $R_{\rm m}$
    dependence at the relevant $kR$ even if we increase $R_{\rm m}$ to $10^{10}$.
The relevant values are therefore not presented because they do not
    meet our definition of ``saturation''.
This, however, is not really a physical issue given that the presented curves already allow us
    to show the existence of optimal axial wavenumbers.
Comparing the solid with the dashed lines, one sees that
    the resistive computations yield a $\tau/P$ that can be considerably larger
    than its TB counterpart.
Take $[l/R, kR] = [0.3, 2]$ for instance, in which case the resistive value
    ($(\tau/P)_{\rm res} = 680$) is an order-of-magnitude larger than the TB one
    ($(\tau/P)_{\rm TB}  = 39$), despite that the TL is still rather narrow.
One also sees that for a given $l/R$, the deviation of the resistive results
    from the TB ones becomes increasingly obvious when $kR$ increases, or equivalently,
    when the axial wavelength $\lambda = 2\pi/k$ decreases.
This is physically understandable because by ``thin-boundary'', the layer width $l$
    should in principle be much smaller than the lengthscales in both the
    transverse (represented by $R$) and longitudinal (here represented by $\lambda$)
    directions.

\begin{figure}
\centering
 \includegraphics[width=1.\columnwidth]{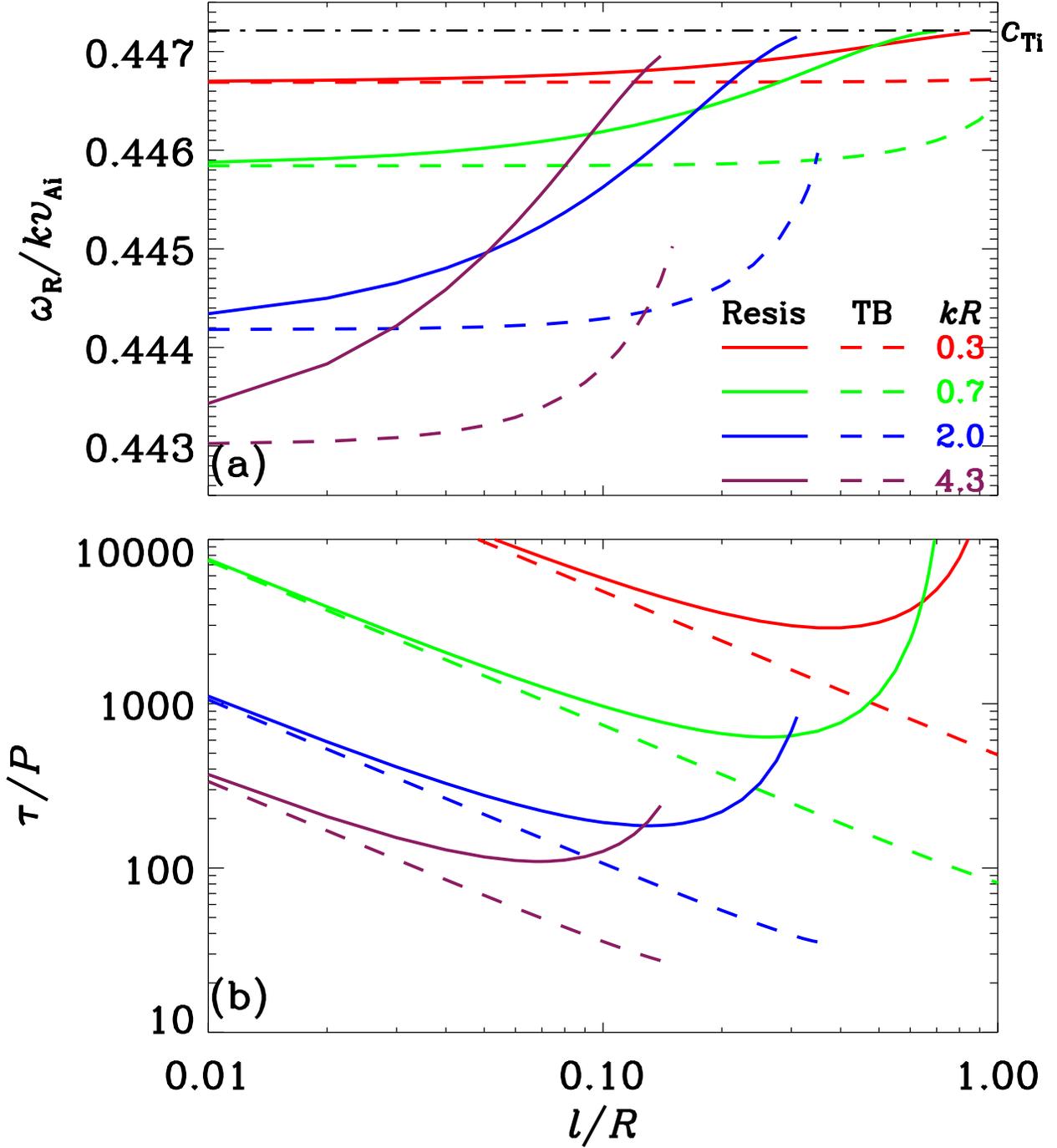}
 \caption{
 Dispersive properties of slow surface sausage modes resonantly coupled to the cusp continuum,
     shown by the dependence on the dimensionless layer width ($l/R$)
    of (a) the axial phase speed ($\omega_{\rm R}/(k v_{\rm Ai})$) and
       (b) the damping-time-to-period ratio ($\tau/P$).
 A number of values of the axial wavenumber $kR$
     are presented by the lines with different colors as labeled.
 In addition to the results from our resistive computations (the solid curves),
    those from the thin-boundary (TB) computations are also shown (dashed).
 }
 \label{fig_dep_l}
\end{figure}

Figure~\ref{fig_dep_l} examines the dispersive properties of SSSMs
    from another perspective, namely, how $\omega_{\rm R}/k$ and $\tau/P$
    depend on the dimensionless TL width $l/R$ for a number of $k$ as labeled.
The results from the resistive computations (the solid curves) are also
    compared with the TB results (dashed).
The comparison between the two sets of curves indicates that in general
    the TB limit overestimates wave damping, and becomes close to the
    resistive values only for relatively small $l/R$.
Define $\epsilon = |(\tau/P)_{\rm res}/(\tau/P)_{\rm TB}-1|$ for convenience.
For a given $kR$, we also define $(l/R)_{\rm TB}$ to be the value beyond which
    $\epsilon$ becomes $\gtrsim 30\%$.
Then one finds that $(l/R)_{\rm TB}$ decreases with $kR$,
    yielding $0.14$ ($0.05$) for $kR = 0.3$ ($2$) for instance.
In other words, $(l/R)_{\rm TB}$ needs to be increasingly small when the axial wavelength decreases,
    as would have been intuitively expected.

\begin{figure}
\centering
 \includegraphics[width=1.\columnwidth]{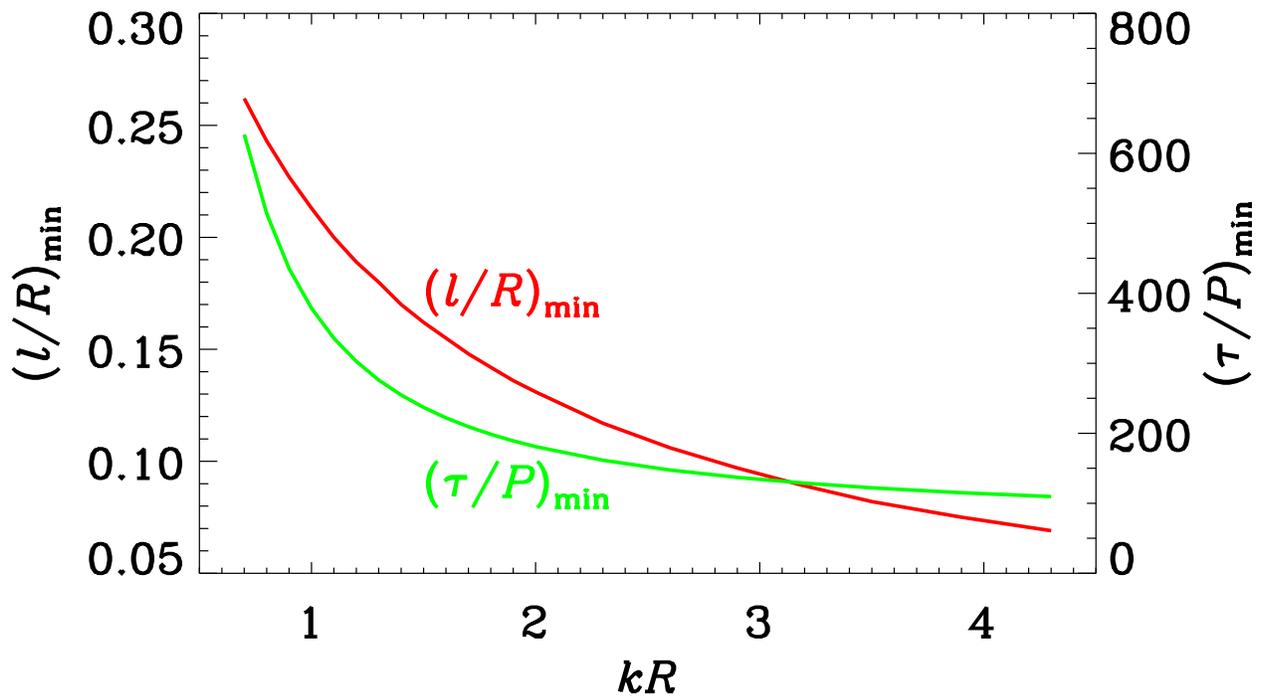}
 \caption{
 Dependence on the axial wavenumber ($k$) of the 
    optimal damping-time-to-period ratio ($(\tau/P)_{\rm min}$)
    and the dimensionless transition layer width ($(l/R)_{\rm min}$)
    where the optimal damping is reached. 
 See text for details.
 }
 \label{fig_lmin_taumin}
\end{figure}

Now examine only the solid curves.
One finds from Figure~\ref{fig_dep_l} that the overall behavior of the axial phase speeds
    and damping-time-to-period ratios does not depend on the value of $kR$.
Examine the blue solid curve in Figure~\ref{fig_dep_l}a ($kR = 2$) for instance.
One sees that $\omega_{\rm R}/k$ tends to increase with $l/R$ and eventually approaches $c_{\rm Ti}$ from below.
We note that the curve terminates at $l/R \sim 0.3$, the reason being not that
    resonant SSSMs are prohibited thereafter.
It is just that the eigen-frequencies do not saturate even at $R_{\rm m}=10^{10}$
    and therefore their values are not presented.
Moving on to Figure~\ref{fig_dep_l}b, one sees that the blue solid curve
    shows a nonmonotonical dependence of $\tau/P$ on $l/R$, and there exists an optimal value of $l/R$
    that maximizes wave damping.
Let $(\tau/P)_{\rm min}$ denote the minimal $\tau/P$ that this curve attains
    and $(l/R)_{\rm min}$ denote where the minimal value is reached.
One sees also that the tendency for $\tau/P$ to increase when $l/R > (l/R)_{\rm min}$
    takes place when $\omega_{\rm R}/k$ becomes sufficiently close to $c_{\rm Ti}$
    (the blue solid curve in Figure~\ref{fig_dep_l}a).
In addition, both $(l/R)_{\rm min}$ and $(\tau/P)_{\rm min}$
    seem to decrease monotonically with $kR$.
This turns out to be indeed the case as shown by Figure~\ref{fig_lmin_taumin},
    where we examine how $(l/R)_{\rm min}$ and $(\tau/P)_{\rm min}$ 
    depend on $kR$ for $kR$ in the range between $0.7$ and $4.3$.
Note that the tendency for $(\tau/P)_{\rm min}$ to decrease together with $(l/R)_{\rm min}$
    when $kR$ increases
    is not to be confused with the tendency for $\tau/P$ to increase with decreasing $l/R$
    when $(l/R) < (l/R)_{\rm min}$ for a given $kR$.
The latter tendency is clear in Figure~\ref{fig_dep_l}b, and is expected given that 
    the cusp resonance will not take place in the extreme case when $l/R = 0$.
What Figure~\ref{fig_lmin_taumin} suggests is that 
    $\tau/P$ is consistently larger than $110$
    for the adopted equilibrium parameters and the wavenumber range.

How do our results connect to the observations of SSSMs by G15?
Actually what leads us to examine the particular range of $kR$ 
    in Figure~\ref{fig_lmin_taumin}
    is based on G15 who examined an SSSM with an axial wavelength $\lambda$ of $4400$~km.
If boldly assuming the same $\lambda$ for SSSMs in photospheric pores in general,
    then one finds that $kR = 2\pi R/\lambda$ is in the range $[0.7, 4.3]$ with $R$
    ranging from $500$ to $3000$~km.
Provided that this assumption holds and that our equilibrium parameters are representative of pores,
    then Figure~\ref{fig_lmin_taumin} indicates that the cusp resonance
    is not efficient in damping SSSMs.
To make Figure~\ref{fig_dep_l} more relevant to G15 where propagating rather than standing SSSMs
    are measured, we also computed the damping-length-to-wavelength ratio ($L_{\rm D}/\lambda$)
    by connecting it to $\tau/P$ through the ratio of the axial group speed ($v_{\rm gr}$)
    to the axial phase speed ($v_{\rm ph}$)~\citep[see e.g.,][Equation~40]{2010A&A...524A..23T}.
This practice is justified because the damping we found is weak.
It turns out that $L_{\rm D}/\lambda$ is nearly identical to $\tau/P$ because $v_{\rm gr}/v_{\rm ph}$
    differs little from unity, which in turn results from the extremely weak dispersion
    (see Figure~\ref{fig_dep_k}a).
The mean radius of the pore that G15 examined is $\sim 1500$~km, and consequently
    $kR \sim 2$.
One reads from the blue solid curve in Figure~\ref{fig_dep_l}b that $\tau/P$
    (and hence $L_{\rm D}/\lambda$) is no smaller than $180$.
We therefore conclude that with our chosen parameters,
    the cusp resonance is unlikely to account for the heavy spatial damping
    of the SSSM in G15, who found a value of $L_{\rm D}/\lambda \sim 1/4$.

Let us come back to Figure~\ref{fig_dep_resis}.
For any given pair of $[l/R, kR]$,
    three distinct intervals of the magnetic Reynolds number $R_{\rm m}$
    show up as far as the behavior of $\tau/P$ is concerned.
As has been discussed in substantial detail, the interval where $\tau/P$ shows no dependence on $R_{\rm m}$
    corresponds to the case where only the cusp resonance operates to damp SSSMs.
This happens for sufficiently large $R_{\rm m}$.
When $R_{\rm m}$ decreases, $\tau/P$ tends to experience an increasingly strong but still gradual
    dependence on $R_{\rm m}$ until a knee occurs
    (see the behavior of $\tau/P$ at $R_{\rm m}\sim 10^{5}$ along the red dotted curve, for instance).
Let this interval be called the intermediate regime.
If further decreasing $R_{\rm m}$ beyond the knees,
    one sees that $\tau/P$ depends linearly on $R_{\rm m}$, as would be expected
    if wave damping is solely determined by resistivity in the case of weak damping.
In this case one expects that $|\omega_{\rm I}| \propto \eta$,
    and consequently $\tau = 1/|\omega_{\rm I}| \propto 1/\eta \propto R_{\rm m}$.
That $\tau/P$ depends also linearly on $R_{\rm m}$ is because of the extremely weak $R_{\rm m}$-dependence of
    $\omega_{\rm R}$ and hence that of $P$.
It is likely that when $R_{\rm m}$ decreases through the knee,
    the resonant SSSM transitions to an ordinary resistive slow mode.
We conjecture this by drawing analogy to resonant kink modes
    in typical coronal loops.
In that case, resonant kink modes cannot be told apart from
    ordinary resistive Alfv\'en modes when the loop boundary becomes sufficiently thick
    (see Fig.~2 in \citeauthor{2007PPCF...49..261V}~\citeyear{2007PPCF...49..261V} for slab computations;
    and Fig.~8 in \citeauthor{2005A&A...441..361A}~\citeyear{2005A&A...441..361A} for cylindrical computations).
However, there is one important difference:
    the SSSMs remain weakly damped in this study, whereas the kink modes are heavily damped 
    in the coronal studies.
It therefore remains open as to why the dependence of $\tau/P$ on resistivity
    becomes qualitatively different when $R_{\rm m}$ crosses the knee. 
We stress that this behavior is not a numerical artifact, because it persists even though we have experimented
    with different choices of grid spacing and
    different trajectories that lead to a given $[l/R, kR, R_{\rm m}]$
    in this three-dimensional parameter space.

Now focus on the red curves, which pertain to $kR = 2$ and therefore are directly relevant to G15.
As mentioned in Section~\ref{sec_intro}, we quote the range of electric conductivity
    as given by \citet[][Table~2]{1983SoPh...84...45K}.
For a pore with radius $R\sim 1500$~km and the adopted equilibrium parameters,
    $R_{\rm m}$ ranges from $3\times 10^{4}$ to $10^7$ as given by the area shaded green.
One sees that the majority of this range lies in the intermediate regime where
    both resistivity and the cusp resonance play a role in damping the SSSM in question.
When $R_{\rm m} = 10^7$, the cusp resonance is nearly the sole factor that damps the SSSM
    for $l/R = 0.1$ (the solid red curve),
    whereas resistivity is still important to some extent
    for $l/R = 0.3$ (dotted).
On the contrary, when $R_{\rm m} = 3\times 10^4$,
    the cusp resonance still plays some role in wave damping for $l/R = 0.1$,
    whereas wave damping is entirely due to resistivity for $l/R = 0.3$.
Regardless, $\tau/P$ at this lower bound is substantially smaller than
    the values due to the cusp resonance alone.
Take $l/R = 0.3$ for instance, for which one finds that $\tau/P$ reads
    $28.9$ when $R_{\rm m} = 3\times 10^4$ but $529$ when $R_{\rm m} = 10^7$.
It is therefore safe to conclude that resistivity should be taken into account
    when the damping of SSSMs in photospheric pores is examined.
However, literally translating $\tau/P$ to $L_{\rm D}/\lambda$, one finds that
    the derived damping-length-to-wavelength ratio is still quite far from the value
    found by G15.

\section{SUMMARY AND CONCLUDING REMARKS}
\label{sec_conclusion}
This study has been motivated by the considerable interest in sausage modes in photospheric
    waveguides, and the spatial damping of propagating
    slow surface sausage modes (SSSMs) as examined by \citet[][G15]{2015ApJ...806..132G} in particular.
Working in the framework of linear resistive MHD, we examined the effects of resistivity and
    resonant coupling to the cusp continuum on the damping of SSSMs in a photospheric
    cylinder, for which the transverse structuring comprises a uniform cord, a uniform exterior,
    and a continuous transition layer (TL) in between.
We found that in general the damping due to the cusp resonance alone is rather weak,
    with the minimal damping-time-to-period ratio (and hence the damping-length-to-wavelength ratio)
    reaching a value of the order $100$.
However, resistivity can be an order-of-magnitude more efficient, making it an indispensable
    ingredient for examining the damping of SSSMs.

Regarding the cusp resonance, the full numerical approach allowed us to examine TLs of arbitrary width,
    and hence allowed a comparison with the TB limit where
    the TL width is assumed to be small.
We found that for the equilibrium parameters we examine, the constraint on
    the range of applicability of the TB approximation
    is rather stringent.
Relevant to G15 for which the dimensionless axial wavenumber $kR \sim 2$,
    we find that the TL width in units of the waveguide radius ($l/R$)
    needs to be $\lesssim 0.05$ for the TB results to be accurate
    within $30\%$.
The TB results tend to overestimate wave damping, by an order-of-magnitude
    when $l/R = 0.3$.

Our study suggests that neither the cusp resonance nor electric resistivity
    can account for the heavy spatial damping of the SSSM measured by G15.
However, this is not the end of the story.
As shown by \citet{2017ApJ...850...44Y}, the damping efficiency due to the cusp resonance
    may be sensitive to the detailed form of the transverse distribution of
    the equilibrium parameters.
On top of that, the values of the equilibrium parameters themselves are expected to
    be important as well.
Our study on the cusp resonance has not exhausted neither of these two options; rather,
    we adopted a particular transverse profile and a given set of equilibrium parameters.
Furthermore, the resistivity we examined only accounted for electron-ion and electron-neutral
    collisions~\citep{1983SoPh...84...45K}.
However, the temperature in the photospheric portion of the pore examined by G15
    may be $\sim 3500$~K, meaning that the ionization degree is likely to be very low.
The Cowling resistivity resulting from ion-neutral collisions may be substantially stronger
    than the Ohmic resistivity we examined.
While the Cowling resistivity has been examined in the wave context for, e.g.,
    kink modes resonantly coupled to the Alfv\'en continuum
    in prominence threads~\citep{2009ApJ...707..662S},
    its effect on the damping of SSSMs in photospheric waveguides
    has yet to be assessed.
{A study along this line of thinking is underway, and seismological applications
    are being explored to infer such parameters as the transverse lengthscales and
    the strengths of both the Ohmic and ambipolar diffusivities. 
}

\acknowledgments
{We thank the referee for his/her constructive comments.}
We thank Tom Van Doorsselaere for carefully reading the first draft.
This work is supported by
    the National Natural Science Foundation of China (41474149, 41604145, 41674172, 11761141002, 41704165),
    and by the Provincial Natural Science Foundation of Shandong via Grant ZR2016DP03 (HY).

\bibliographystyle{apj}
\bibliography{saus_cusp}

\IfFileExists{\jobname.bbl}{} {\typeout{}
\typeout{****************************************************}
\typeout{****************************************************}
\typeout{** Please run "bibtex \jobname" to obtain} \typeout{**
the bibliography and then re-run LaTeX} \typeout{** twice to fix
the references !}
\typeout{****************************************************}
\typeout{****************************************************}
\typeout{}}

\clearpage

\clearpage

\clearpage

\clearpage

\clearpage

\end{document}